\documentclass[conference]{IEEEtran}
\IEEEoverridecommandlockouts
\usepackage{cite}
\usepackage{amsmath,amssymb,amsfonts}
\usepackage{algorithmic}
\usepackage{graphicx}
\usepackage{textcomp}
\makeatletter

\newcommand{\Rmnum}[1]{\expandafter\@slowromancap\romannumeral #1@}
\makeatother
\usepackage{multirow, makecell}
\def\BigRoman{\uppercase\expandafter{\romannumeral\number\count 255 }}
\def\Romannumeral{\afterassignment\BigRoman\count255=}
\usepackage{tikz}
\def\BibTeX{{\rm B\kern-.05em{\sc i\kern-.025em b}\kern-.08em
    T\kern-.1667em\lower.7ex\hbox{E}\kern-.125emX}}
  
\begin{document}

\title{Motor Imagery Classification of Single-Arm Tasks Using Convolutional Neural Network based on Feature Refining
\footnote{{\thanks{\hrule Research was partly supported by Institute of Information \& Communications Technology Planning \& Evaluation (IITP) grant funded by the Korea government (No. 2017-0-00432, Development of Non-Invasive Integrated BCI SW Platform to Control Home Appliances and External Devices by User’s Thought via AR/VR Interface) and partly funded by Institute of Information \& Communications Technology Planning \& Evaluation (IITP) grant funded by the Korea government (No. 2017-0-00451, Development of BCI based Brain and Cognitive Computing Technology for Recognizing User’s Intentions using Deep Learning).}
}}
}

\author{\IEEEauthorblockN{Byeong-Hoo Lee$^1$, Ji-Hoon Jeong$^1$, Kyung-Hwan Shim$^1$, Dong-Joo Kim$^1$}
\IEEEauthorblockA{{$^1$Department of Brain and Cognitive Engineering, Korea University, Seoul, Republic of Korea}}

{bh$\_$lee@korea.ac.kr, jh$\_$jeong@korea.ac.kr, kh$\_$shim@korea.ac.kr, dongjookim@korea.ac.kr}
}


\maketitle

\begin{abstract}
Brain-computer interface (BCI) decodes brain signals to understand user intention and status. Because of its simple and safe data acquisition process, electroencephalogram (EEG) is commonly used in non-invasive BCI. One of EEG paradigms, motor imagery (MI) is commonly used for recovery or rehabilitation of motor functions due to its signal origin. However, the EEG signals are an oscillatory and non-stationary signal that makes it difficult to collect and classify MI accurately. In this study, we proposed a band-power feature refining convolutional neural network (BFR-CNN) which is composed of two convolution blocks to achieve high classification accuracy. We collected EEG signals to create MI dataset contained the movement imagination of a single-arm. The proposed model outperforms conventional approaches in 4-class MI tasks classification. Hence, we demonstrate that the decoding of user intention is possible by using only EEG signals with robust performance using BFR-CNN.\\
\end{abstract}

\begin{small}
\textbf{\textit{Keywords-brain-computer interface; electroencephalogram; motor imagery; convoulutional neural network}}\\
\end{small}

\section{Introduction}
Brain-computer interface (BCI) decodes brain signals to understand user intention and status that can be used for external device control. Since brain signals contain diverse information about user status, many studies have attempted to understand brain signals through BCI \cite{wolpaw2002brain,C3,ECoG,MRCP,B1}. Invasive BCI directly places the electrodes on the brain to acquire high-quality brain signals such as electrocorticogram (ECoG) \cite{ECoG2}. However, there are many safety issues associated with invasive BCI because it involves surgery to implant electrodes. On the other hand, non-invasive BCI uses electroencephalogram (EEG) because it is easy to acquire without brain surgery. EEG-based BCI has several paradigms for signal acquisition such as motor imagery (MI) \cite{A2,C2,kam}, movement-related cortical potential (MRCP) \cite{MRCP}, and event-related potential (ERP) \cite{ERP,EEG,A1}. As applications of EEG-based BCI, speller \cite{speller} and wheelchair \cite{wheelchair}, and drone \cite{drone} were commonly used for communication between user and devices. Among these paradigms, MI is related to specific potentials from the supplementary motor area and pre-motor cortex \cite{MI}. When the user imagines specific movements, event-related desynchronization/synchronization (ERD/ERS) patterns are generated in supplementary motor area and pre-motor cortex \cite{ERD}. MI paradigm captures these patterns to detect user intention. Due to its origin, MI is commonly used for recovery or rehabilitation of the user's motor functions using external devices \cite{roboticarm}. Additionally, MI-based BCI provides extra motor functions using robotic arm \cite{C3}.

\begin{figure*}[t]
\centerline{\includegraphics{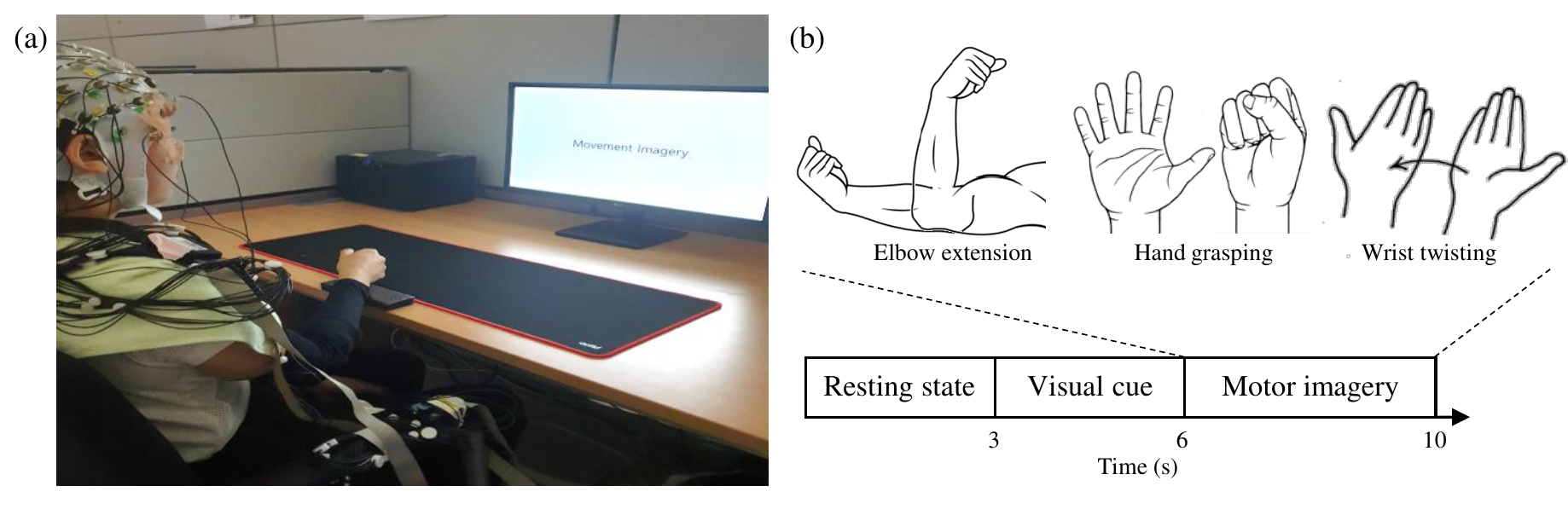}}
\caption{(a) Experimental environment for EEG data acquisition. (b) Experimental paradigm of single-arm tasks. From 0 to 3 seconds, resting state was given as relaxation. After resting state, 3 seconds of visual cue like above figures was given for readiness. Finally, 4 seconds of imagery period was given.}
\label{fig}
\end{figure*}

EEG is an oscillatory and non-stationary signal thus decoding EEG signals is challenging work \cite{nonstationary,B2}. Similar to the denoising technique in computer vision \cite{book,A3}, EEG signal should be treated after denoising using filters. A number of MI classification methods have been developed to achieve satisfactory classification performance. Filter bank common spatial patterns (FBCSP) is conventional feature extraction method to decode EEG signal using spectral power modulations \cite{FBCSP}. Linear discriminant analysis (LDA) is jointly used with FBCSP as a classifier. Cho et al. \cite{channel} used FBCSP with regularized linear discriminant analysis (RLDA) to decode MI tasks focusing on a single category of MI tasks such as hand grasping and arm reaching. Convolutional neural network (CNN) approaches are applied in BCI \cite{CNN}. Schirrmeister et al.\cite{deepconvnet} proposed three different types of CNN-based models depending on the number of layers, inspired by FBCSP. Among the three models, ShallowConvNet extracts log band power features. MI classification performance of the ShallowConvNet is better than the DeepConvNet which is designed for general purpose dealing with signal amplitude. Using the depth wise and separable convolutions, CNN performs classification well regardless of the types of EEG signals including MI \cite{EEGNET}. However, these studies mainly focused on simple tasks using competition dataset (left-hand, right-hand, foot, and tongue) and classes are not related to each other to perform sequential work such as drinking water and opening the door. Since the commands are not intuitive, artificial command matching should be required to control external devices.

In this study, we collected three different types of MI tasks of a single-arm: elbow extension, wrist-twisting, and hand grasping to perform sequential upper limb works. Second, we proposed a band-power feature refining convolutional neural network (BFR-CNN) which has only two convolution blocks for MI classification by extracting band-power features. It is designed to classify single-arm MI tasks without artificial command matching. Finally, the proposed BFR-CNN achieved robust classification performance in the 4-class single-arm MI tasks classification. 

This paper is structured as follows. Section {\Romannumeral 2} gives a description of the data acquisition, dataset for evaluation, and the proposed BFR-CNN model. Section {\Romannumeral 3} presents the results of classification accuracies, performance comparison using other models and discusses the advantages and limitations. In session {\Romannumeral 4}, conclusions and future work are described.\\

\begin{figure}[t]
\centerline{\includegraphics[width = \columnwidth]{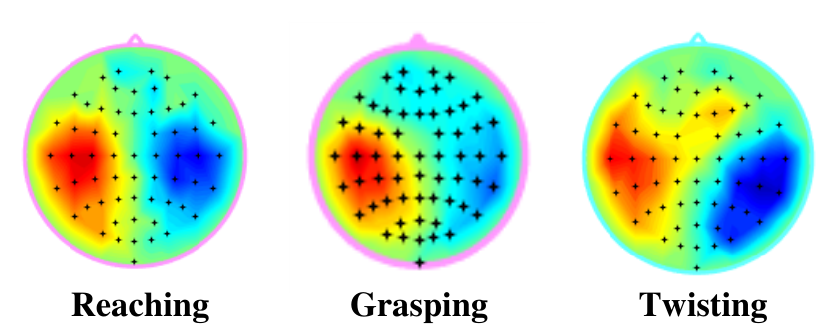}}
\caption{Representative topoplots of each MI task. 8-12 Hz of frequency band was selected from a subject. High amplitudes were observed in the left side of supplementary motor area and the pre-motor cortex because the subject is right-handed.}
\label{fig}
\end{figure}

\section {Methods}
\subsection{Data description}
Data acquisition process was conducted with eight healthy subjects at the age of 22-30 (6 right-handed males and 2 right-handed females). We used EEG signal amplifier (BrainAmp, BrainProduct GmbH, Germany) to record EEG signals. The sampling rate was 1,000 Hz and a band-pass filter (1-60 Hz) was applied in all channels. We applied 60 Hz notch filter to remove noise from the wires. Brain Products VisionRecorder (BrainProduct GmbH, Germany) recorded and filtered raw EEG data from the subjects. 64 Ag/AgCl electrodes in 10-20 international system were used. The FPz and FCz channels were selected as ground and reference respectively. Impedance of each electrode was measured to maintain the impedance below 10k$\Omega$ using conductive gel. 64 EEG channels were used for data acquisition and we selected 24 channels (F3, F1, Fz, F2, F4, FC3, FC1, FC2, FC4, C3, C1, Cz, C2, C4, CP3, CP1, CPz, CP2, CP4, P3, P1, Pz, P2, and P4) for evaluation \cite{channel}. These channels are placed on the somatosensory area and pre-motor cortex. During the data acquisition experiment, every subject performed the 150 trials of MI tasks (i.e., 50 trials of elbow extension, twisting and grasping tasks). Relaxation was given before the imagery period and extracted as a resting state (Fig. 1). Subjects were asked to imagine specific muscle movements. Collected MI dataset were resampled at 250 Hz for the classification and it contained 3 classes of single-arm tasks and resting state. Data validation was conducted using FBCSP algorithm and RLDA for each MI task. The protocols and environments were reviewed and approved by the Institutional Review Board at Korea University [1040548-KU-IRB-17-172-A-2].

\begin{figure*}[t!]
\centerline{\includegraphics[scale= 0.9] {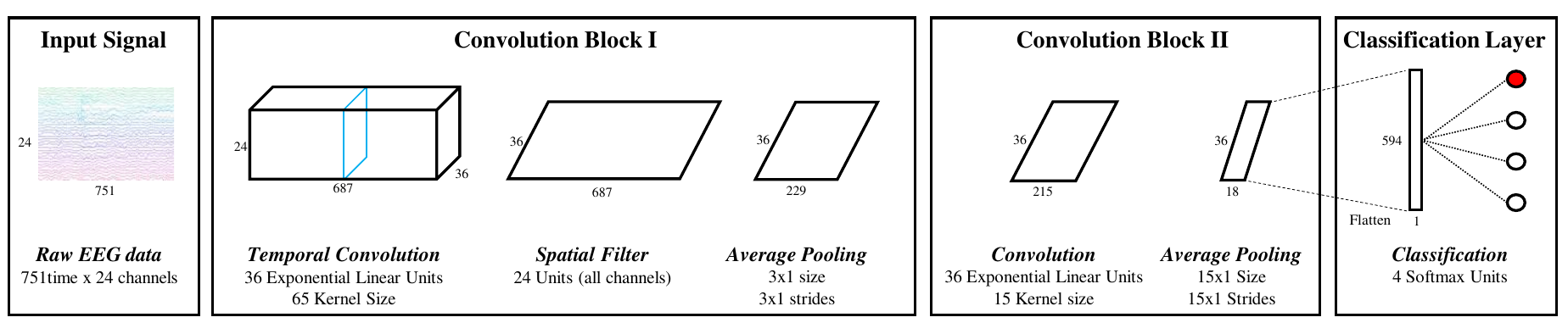}}
\caption{Overall flowchart of the proposed BFR-CNN. It consists of two convolution blocks. The first convolution block was designed for creating a receptive field and the second block is for feature refining.}
\label{fig}
\end{figure*}
\begin{table}[!t]
\caption{Comparison of MI tasks classification results}
\resizebox{\columnwidth}{!}{%
\renewcommand{\arraystretch}{1.4}
\begin{tabular}{cccccc}\hline
     & \textit{BFR-CNN} & \textit{DeepConvNet} & \textit{ShallowConvNet} & \textit{EEGNet} & \textit{FBCSP+RLDA} \\ \hline
sub1 & 0.82 & 0.74        & 0.83           & 0.68   & 0.68      \\ 
sub2 & 0.83 & 0.61        & 0.74           & 0.63  & 0.70       \\ 
sub3 & 0.84 & 0.78        & 0.83           & 0.84   & 0.69       \\
sub4 & 0.80 & 0.50        & 0.72           & 0.58   & 0.64       \\
sub5 & 0.90 & 0.71        & 0.85           & 0.71   & 0.75       \\
sub6 & 0.80 & 0.59        & 0.71           & 0.63   & 0.53       \\
sub7 & 0.84 & 0.60        & 0.76           & 0.66   & 0.65       \\
sub8 & 0.88 & 0.55        & 0.68           & 0.60   & 0.72       \\\hline
\textbf{Avg.} & 0.84 & 0.64        & 0.77           & 0.67   & 0.67       \\
Std. & 0.04 & 0.10        & 0.06           & 0.10   & 0.11     \\  \hline
\end{tabular}
}
\end{table}

\begin{figure}[t!]
\centerline{\includegraphics[width = \columnwidth]{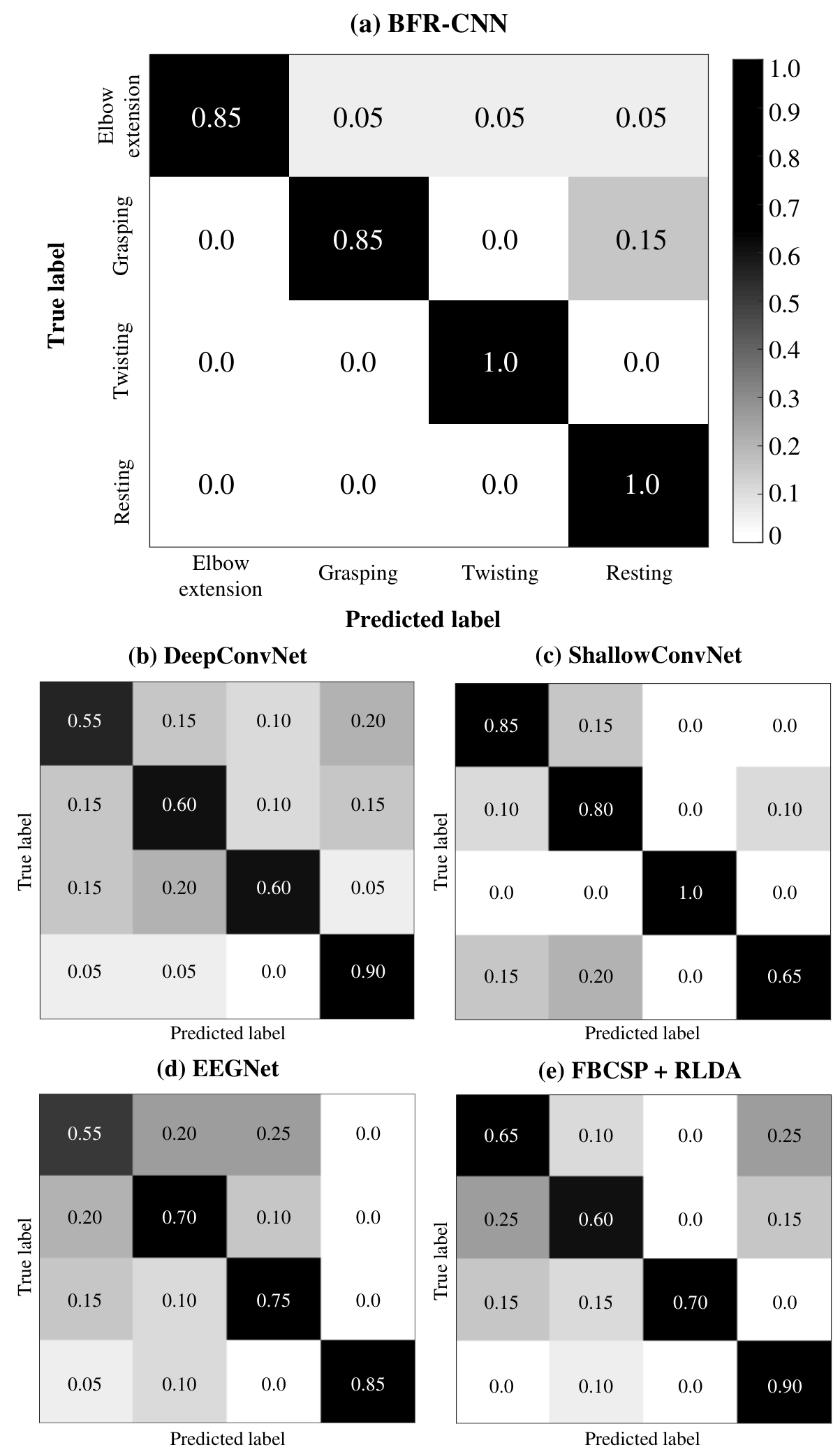}}
\caption{Representative confusion matrices of each model. Through these confusion matrices, we could analyze the classification tendencies of each model.}
\label{fig}
\end{figure}

\subsection{BFR-CNN}
BFR-CNN is a singular CNN architecture and it is designed for single-arm MI tasks classification. Raw EEG signal contains large amounts of information (channel by time size of matrix) which are not relevant to MI tasks. Therefore, if the classification model is able to extract features and refine them into more relevant features, then the classification performance can be improved. Because the classes of our dataset are composed of single-arm tasks, we assumed that spatial features from the restricted cortex region would not be sufficient to be used on CNN. In addition, since the EEG signals have a high temporal resolution, a higher performance can be achieved by extracting frequency features rather than spatial features \cite{highertemporal}. We were inspired by the concept of shallowConvnet which extracts log-band power features. Considering class complexity of our data, we assumed that using more and refined features would be proper. We conducted frequency domain analysis and there was high amplitude in the similar brain regions (left side of somatosensory cortex and motor cortex) found in topoplot of each MI task (Fig. 2). Thus, we attempted to develop the shallow CNN architecture that would extract frequency features that are highly relevant to single-arm MI tasks through the convolutional layer. The first convolution block consists of a temporal convolution layer, spatial filter layer, and average pooling layer \cite{averagepooling}. Spatial filter was applied along the input channels to reduce the dimensionality as a single input channel. We set the temporal filter size to a quarter of sampling rate to remove the ocular artifact creating a receptive field above 4 Hz. The second block was designed to refine band-power features. We comprised the second convolution block with convolution layer and average pooling layer to reduce the number of features that are less relevant for classification. The last layer contains softmax function with the flatten layer for classification which normalizes output probability distribution. The exponential linear unit (ELU) was applied as an activation function in every convolution block \cite{ELU}. We used adam optimizer \cite{ADAM} and cross-entropy loss function for training \cite{entropy}. The overall flowchart of BFR-CNN is described in Fig. 3.\\

\section {Experimental results and discussion}

For the evaluation, we set mini-batch size as 32 and 200 times of training epochs. Evaluation environment was Window 10 desktop with specification Intel(R) Core i7-7700 CPU at 3.60 GHz, 32GB RAM, and Geforce Titan XP GPU. All comparisons were conducted under the same conditions.

Table \Rmnum{1} shows a comparison results of classification. The average accuracy of the BFR-CNN is 0.84 as the highest accuracy among the comparison groups. However, the ShallowConvNet ranked as second-place records 0.77 and that is because it extracts log band power features similarly BFR-CNN which refines band-power features. The remaining methods show similar classification performance. The DeepConvNet records the lowest performance that is because it is designed for general purpose especially concerning signal amplitude. EEGNet is also designed to decode EEG signal regardless of its dominant features even in MI classification thus EEGNet classifies slightly better than DeepConvNet. Interesting thing is that FBCSP with RLDA performs MI classification as well as EEGNet even it is not a deep learning. Through the comparison, we confirm that using the band-power features is advantageous for MI task classification, and refinement can yield higher classification performance.

Fig. 4 is the confusion matrices of classification results. DeepConvNet tends to confuse all MI tasks (elbow extension, grasping and twisting) especially elbow extension but it classifies relatively well the resting state. The ShallowConvNet clearly classifies twisting but confuses elbow extension, grasping and resting state. Unlike other methods, ShallowConvNet is weak in classifying resting states. On the other hand, the ShallowConvNet performs MI tasks classification with high accuracies. EEGNet strongly confuses the elbow extension class with the grasping and twisting. However, none of the MI tasks have been misclassified as resting state. FBCSP with RLDA classifies MI tasks as well as EEGNet but it shows higher classification accuracy in elbow extension classification. BFR-CNN clearly classifies twisting and resting state. Like ShallowConvNet, there is a tendency to slightly confuse elbow extension and grasping. Overall, we find that all methods used in this study tend to confuse MI tasks rather than resting state.\\

\section{Conclusion and Future works}
\label{sec:print}

In this paper, we propsed a BFR-CNN that refines band-power features to classify single-arm MI tasks. The decoding of MI dataset is time-consuming and costly work because it is oscillatory and non-stationary signals. To improve MI classification performance, we proposed BFR-CNN to extract and refine frequency features that are highly relevant to MI. Through the evaluation, we demonstrated that the BFR-CNN achieved the highest classification accuracies compared to existing approaches. Thus, the proposed model can be applied to control external devices with high performance such as a robotic arm. 


\section{Acknowledgement}
The authors thanks to J.-H. Cho for their help with the dataset construction and  discussion of the data analysis.\\

\bibliographystyle{IEEEbib}
\bibliography{refs}

\end{document}